# A Novel Poisoned Water Detection Method Using Smartphone Embedded Wi-Fi Technology and Machine Learning Algorithms

**Halgurd S. Maghdid[1,*], Sheerko R. Hma Salah[1], Akar T. Hawre[1], Hassan M. Bayram[1], Azhin T. Sabir[1], Kosrat N. Kaka[3], Salam Ghafour Taher[3], Ladeh S. Abdulrahman[1], Abdulbasit K. Al-Talabani[1], Safar M. Asaad[1], Aras Asaad[1,2]**

[1] Department of Software Engineering, Faculty of Engineering, Koya University, Kurdistan Region of Iraq
[1,2] School of Computing, The University of Buckingham, UK
[3] Department of Chemistry, Faculty of Science and health, Koya University, Kurdistan Region of Iraq

**Abstract**: Water is a necessary fluid to the human body and automatic checking of its quality and cleanness is an ongoing area of research. One such approach is to present the liquid to various types of signals and make the amount of signal attenuation an indication of the liquid category. In this article, we have utilized the Wi-Fi signal to distinguish clean water from poisoned water via training different machine learning algorithms. The Wi-Fi access points (WAPs) signal is acquired via equivalent smartphone-embedded Wi-Fi chipsets, and then Channel-State-Information CSI measures are extracted and converted into feature vectors to be used as input for machine learning classification algorithms. The measured amplitude and phase of the CSI data are selected as input features into four classifiers k-NN, SVM, LSTM, and Ensemble. The experimental results show that the model is adequate to differentiate poison water from clean water with a classification accuracy of 89% when LSTM is applied, while 92% classification accuracy is achieved when the AdaBoost-Ensemble classifier is applied.

*** Corresponding author**: email: halgurd.maghdid@koyauniversity.org

## 1- Introduction:

Healthcare is one of the most important sectors in our daily life that requires certainty and regular checking [1]. Current healthcare systems help to conduct diagnosis and reduce the medical treatment. Despite the need for a good financial support, sometimes diagnosis and cure may take a period of time due to having various diseases for different reasons such as food, drink, environment, etc. Water drinking flows through the organs of human body, as it is known that it makes up a significant fraction of the body [2]. For example, healthy humans with regular activities need around 2.7 litre to 3.7 litre daily water intake [3]. Further, water is one of the essential fluids for humane body, it's not just for quenching the thirst, it is also important to hydrate the body and keeping the body organ functions work properly and healthy. However, for whatever reason, sometimes people deliberately put poison in clean water, making it poisonous, or sometimes the water is spontaneously poisoned by chemicals elements (such as toxic element [4]) in the water sources. Furthermore, the cumulated poisoning mainly comes from water, as human drink water frequently, and even trace amounts of toxins in water cumulates with in short period of time and cause serious health problems and in serious cases cause death [5].

There are several hidden toxic chemicals soluble in water with no colour, flavour and some are reported to cause acute poisoning [6]. Among toxic compounds in the list are metals (Pb, Cd, Cr, Ni, and Zn), particularly heavy metals. Lead (Pb) harmfulness is significant consequences for the human body, and has a lethal dose around (LD50 93 mg/kg Chronic Toxicity) to guarantee death by ingestion. Lead poisoning happens when its concentration builds up in the body by entering repeated doses from water or any sources, and often over months or years [7]. This seriously affect mental and physical development with enormous numbers of disease (anaemia, high blood pressure, and chronic nervous system disorder), and can cause coma, seizures and death in very high levels [8]. The research in [9] suggests that lead nitrate [$Pb(NO_3)_2$] exposure lead to the activation of several cellular and molecular processes including, induction of cell death, externalization of phosphatidylserine, and activation of caspase-3 in human leukemia (HL-60) cells.

Equally, in the last decades the growth of technology marked an important milestone in the quality of life, including health, education, economy, and industry. It even reflected on the individual's life, whether they are aware of it or not. In addition to that, a numerous number of smart devices such as smartphones, wearable devices, personal device assist (PDA), and tablets are witnesses on an immense demand on emerged technologies [10]. Hence, several methods and techniques have been proposed to detect such kind of poisoned water using different types of technology including: chemical instruments in laboratory [11], using optical sensors [12], using pH sensors [13], and using conductivity sensors [14]. However, all these solutions have its own limitations which are centred on: either suffering from users' intervention, incurs huge cost, time consuming, or using extra (sometime unnecessary) sensors. To address these issues and limitations, this study proposes a novel method to detect poisoned water using off-the-shelf Wi-Fi chipset within a short time. Specifically, these are embedded on the smartphones as well as running a trained AI model. The proposal also provides low-cost solution without user interaction with the water. Therefore, the **contribution** of this study is to:

- Provide a novel method to detect poisoned water using smartphone embedded Wi-Fi technology and artificial learning model without user intervention, within short time, and without extra sensors (i.e. low cost solution).
- Building a database of Wi-Fi CSI data (specifically, amplitude and phase) from clean and poisoned water which will be public for researchers for future investigations.

The rest of the article is organised as follows. In the section 2, the current poisoned and water contamination detection methods and techniques are presented and discussed. This is followed by presenting proposed approach of using Wi-Fi signals and smartphone CSI, materials, detection process, and data collections in section 3. While section 4 provides the experimental results and discussion of obtained results. Finally, section 5 concludes on the proposal as well as presents the next steps and challenges of the proposed approach.

## 2- Related Work:

Smartphones and tablets, equipped with computational power, sensors, and networking capabilities make them suitable as measuring devices. There is a long history of utilizing mobile phone sensing and data visualization features through research and commercially produced accessories. Zhang et. al [15] developed a method to utilise wireless signals to categorise drinks into six groups of soft drinks, fruit and vegetable juice, energy drinks, tea drinks, milk beverages and coffee drinks. They reported on average 87% accuracy detection for each of the drinking categories. The use of radio frequency identification (RFID) to learn the quality of food has been explored in [16]. More specifically, they exploit electromagnetic interactions between wireless stickers placed on food containers and materials inside to determine food quality and safety in which they obtained promising results when tested on 5 classes of adulteration, see section 4.2. In the same vein, RFID used to differentiate 10 types of drinks and 6 types of solid targets using phase and received Signal Strength (RSS) measurements. In [17], the authors proposed to create a biosensor that attaches to a smartphone, utilizing bacteria to detect unsafe levels of arsenic in water. To detect the level of contamination in water, authors in [17] used the smartphone to generate clear and easily understandable patterns, similar to volume bars. The researchers have emphasized the urgent need for cost-effective and on-site solutions for contaminated water sources, particularly in resource-limited countries where there is a scarcity of trained personnel and healthcare facilities for water testing. They suggest that this new device can replace the current testing methods, which are complex to operate, require specialized laboratory equipment and can produce harmful chemicals. Furthermore, touch screen of mobile phones used by researchers from Cambridge university in England to measure liquid electrolyte conductivity for sensing drinking water contamination and soil health [18]. Hojris et al. in [19] developed an optical, on-line bacteria sensor to generate 3D images and then count suspended particles to differentiate bacteria from abiotic particles. To test the feasibility of Ultra-wideband (UWB) wireless signals, Dhekne et al. [20] proposed an algorithm to use UWB to identify 33 types of liquid without using any feature extraction of machine learning model. Instead, they used principles of physics in which they estimated the primitively of the liquid to model the behaviour of radio signals inside liquids with 9% error rate. On the other hand, there are use cases of using buoys equipped with different type of Ponsel sensors (Ponsel PHEHT, Ponsel OPTOD and Ponsel C4E) placed on the surface of river water to measure the quality of water and transmits measured data wirelessly to gateways and then to a server to generate alerts in case of water pollutions [21].

Previous studies for material identification require either touching the material physically, expensive or time consuming. In this study, we propose the use of smartphone Wi-Fi signal technology to detect water intoxication in deferent levels.

## 3. The Proposed Approach

Quality of drinking water and even contaminated water issue still has a dangerous effective on human being life in most of the developing countries. Even in developed countries the issue of poisoned water, specifically for the targeted very important persons (VIP) or ordinary persons, is an issue in the security scenario. A large number of research and experiments have been conducted to detect and predict the quality level of drinking water. Equally, current technologies including wireless sensing have real impact on improving our daily life quality. In addition to that, smart handheld devices such as wearable devices, smartphones, and tablets are embedded with such technologies. The smartphones' capabilities in terms of hardware and software can manage and perform very well to tackle most of the aforementioned health issues. Therefore, in this study, a novel approach is proposed to detect the poisoned water via using on-board technologies and smart software model without the need of inventing new sensors.

### 3.1 Materials and Methods

The current wireless sensing involves the use of universal wireless signals, such as electromagnetic, light, and sound waves, to gather information from people and their surroundings without direct contact. This technology has numerous potential applications in fields such as the Internet of Things, AI, healthcare, and defence [22].

Furthermore, the basic idea behind wireless sensing is that the transmitter sends out a signal, which encounters physical occurrences such as direct reflection and scattering during its transmission, creating multiple pathways for the signal to travel through [23]. The signal received at the receiver side through multipath transmission which carries information about the signal's propagation environment. By analysing the variations of wireless signals during transmission, wireless sensing technology can determine the characteristics of the signal's propagation environment, allowing for scene sensing to occur [24]. Hence, such sensing technology doesn't need any user interaction or doesn't need to install extra sensors to sense the environment changes or people activities.

One of the most deployed wireless sensing is the Wi-Fi technology [25]. This is because the Wi-Fi signals are already available around us for Internet services and networking. Therefore, wireless sensing using Wi-Fi technology (Wi-Fi Sensing) provides low-cost solutions and without user intervention.

In another vein, to sense the environmental changing through Wi-Fi sensing, Wi-Fi signal measurements are needed. The most effective Wi-Fi signal measurements are the channel-state-information (CSI) and received-signal-strength (RSS) [26]. However, the RSS has the multipath propagation issue and will not be accurate for the sensing process. Therefore, in this work we opt to use CSI measurements for the purpose of poisoned wated detection.

The CSI explains the process by which Wi-Fi signals travel from the source to the receiver and the effects of such propagation, including scattering and attenuation. It also sheds light on how the physical surroundings affect wireless signals. With the CSI, detailed information can be obtained about the propagation of Wi-Fi signals, including time delays, reduction in amplitude, and phase changes, which can provide valuable insights into signal propagation. In comparison to RSS, CSI provides a more precise and accurate reading, that is why this study opt to choose the CSI data to design a novel approach for detecting the poisoned water [26].

The CSI data of the received Wi-Fi Access Point (WAP) signals can be measured at the physical layer. However, the CSI data of the received signals can't be retrieved through the on-board smartphones Wi-Fi chipsets, since this functionality is not enabled at the physical layer by default. This is because enabling Wi-Fi technology functionalities are different when it is embedded on the mobile platform and sometimes because of the commercial issues which has been standardized by different companies. Fortunately, authors in [26] patched the firmware of the Wi-Fi chipset on the Nexus smartphones. However, we could not use the Nexus smartphone, since such smartphones are not available in the Iraqi Market. Due to this limitation, initially, in this study, two Wi-Fi ESP32 chipsets are used, the first one represents a Wi-Fi transmitter and it is configured as a Wi-Fi access point, to advertise beacon frames every 100 milliseconds. The second chipset represents a Wi-Fi receiver (i.e. representing as a smartphone Wi-Fi chipset), which is configured as monitor mode to receive the WAP's signal. Note that the measured CSI data on Wi-Fi ESP32 is already enabled and it can be retrieved [27] similar to the firmware used in [26], which is the same type of Wi-Fi receiver as the real Nexus smartphone firmware.

### 3.2 Detection Process

The complete water poisoning detection process using Wi-Fi signal measurements and machine algorithms is shown in Figure 1.

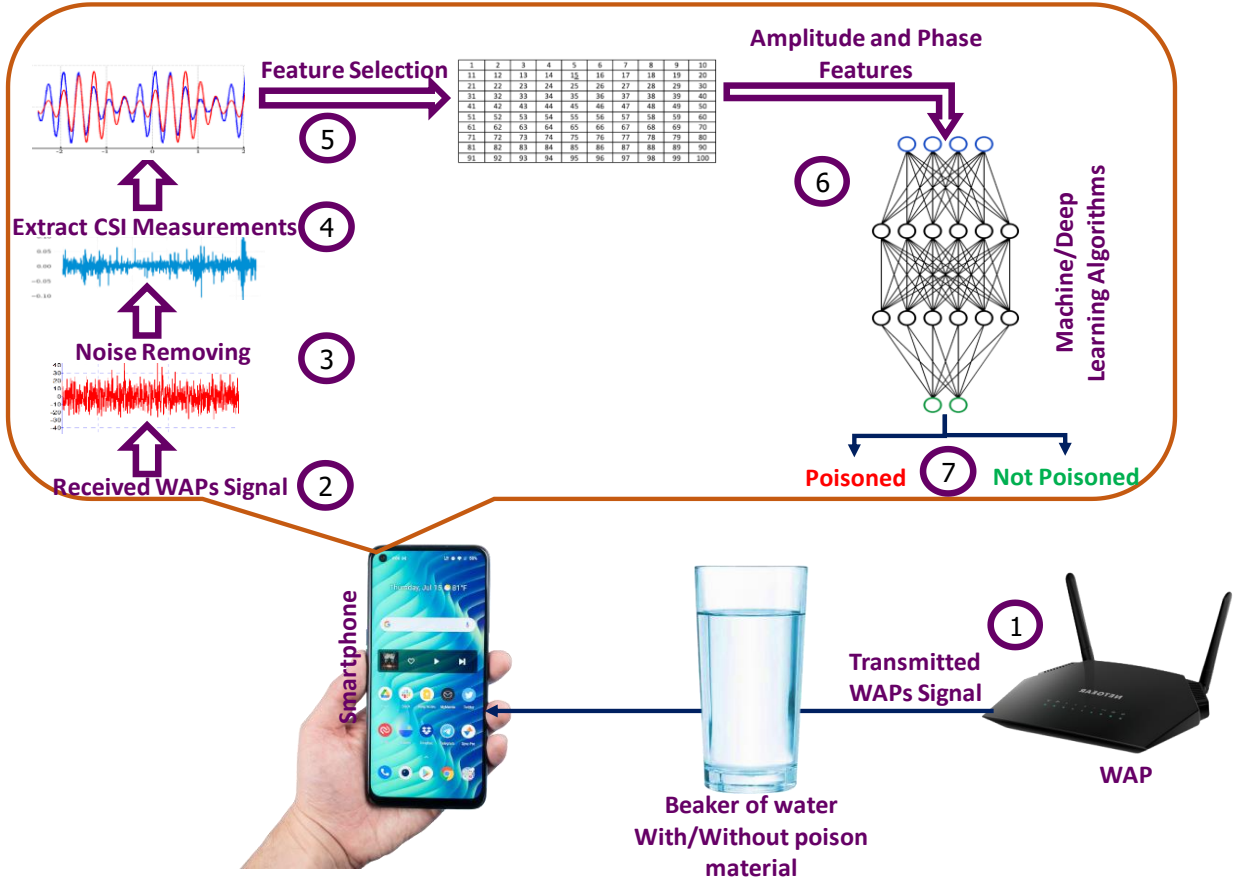

Figure 1. The general process of the poisoned water detection using Wi-Fi signals and Machine Learning algorithms.

The detection process starts by transmitting the **Wi-Fi signals via the WAP**. Then the WAP signals are propagated and penetrated through a **glass (beaker) of water**, which clean or poisonous water. Next, the **smartphone** (here, the Wi-Fi chipset) constantly receives the penetrated WAP signals and extracts the WAP's signal measurements. However, due to the environmental and chipset conditions, the received WAP's signal should be cleaned from the **noise** including the multipath signals, un-targeted WAP signals (i.e. the Wi-Fi signals from other WAPs in the vicinity), CSI interference in both time and frequency domains and dynamic power transmission.

Once the noise is removed, then the **CSI data** could be calculated from the WAPs signal measurements. The WAPs signal measurements includes: CSI data, signal mode, RSS, MAC address, channel number, and signal length. However, in this work we only utilize the CSI data to generate the CSI features. Therefore, the next step is to extract the intended features from the CSI data. The CSI data can provide different features including: kurtosis of CSI, frequency of CSI, impulse factors, time domain energy, clearance factor, peak values, amplitude, and phase measurements. Although, all these aforementioned features are generated during the data collections, however, this study is relying only on the **amplitude and phase** measurements, since they are provided good sensing data to get better detection accuracy.

The CSI data of the received WAP signal can be expressed as follows:

$$\vec{Y} = H \cdot \vec{N} + \vec{X} \qquad (1)$$

where $\vec{Y}$ and $\vec{X}$ are representing the transmitting and receiving signal vectors, the $H$ matrix denotes the channel state information, while the $\vec{N}$ vector represents additive white Gaussian noise. However, the CSI data is a collection of channel information for each subcarrier, and it can be estimated using $\vec{Y}$ and $\vec{X}$. To further understand the H matrix is expressed in equation (2):

$$H = [H_1 \; H_2 \; H_3 \; ... \; H_n]^T \qquad (2)$$

where the *n* is the number of subcarriers, and the $H_i$ appears in the plural form, as expressed in equation (3):

$$H_i = |H_i|\exp\{j \angle H_i\} \qquad (3)$$

where $|H_i|$ and $j \angle H_i$ are the amplitude and phase measurements of the subcarrier *i*.

After the process of calculating the amplitude and phase measurements of all the subcarriers, the measurements are concatenated all together as a row vector to be used as input to the machine learning algorithms. Finally, different machine learning and deep learning algorithms, including Support-Vector-Machine, k-nearest neighbour (k-NN), long-short-term memory (LSTM) and AdaBoost-Ensemble learning algorithm are deployed to classify whether the water is poisonous or not.

### 3.3 Dataset Collection

In this study, a new dataset for the WAP signals is collected by penetrating the signal through a glass of water in a chemistry laboratory at Faculty of Science and Health in Koya University. Two Wi-Fi ESP32 chipsets are used, the models of the chipsets are shown in Figure 2, one is working as a WAP to transmit the Wi-Fi signals, while the second one is working as a Wi-Fi receiver to receive the WAPs signals. The two chipsets are installed on a table, when the glass of the water is between the two chipsets within 50 centimetres far from each chipset, as shown in Figure 3.

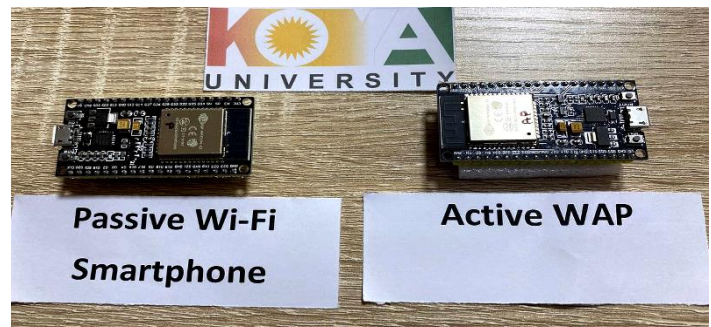

Figure 2: The two Wi-Fi EPS32 chipsets, left one is representing as a Smartphone, right one is working as WAP.

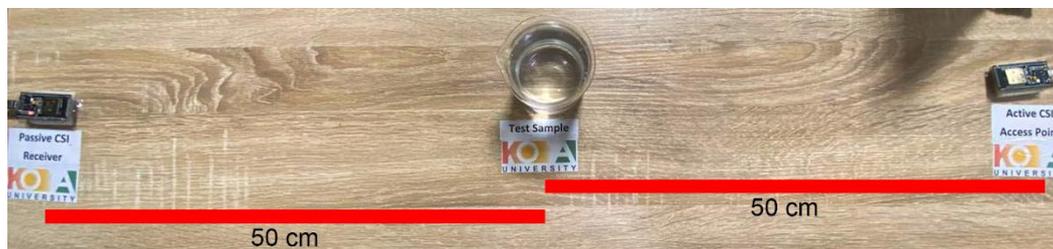

Figure 3: A snapshot of the testbed materials including: the two Wi-Fi chipsets and the glass of water between them.

The data collected within three trials. With the first trial, when the glass is filled with clean water, 2644 CSI measurements from WAP signals are recorded. While in the second and third trials, when the glass is filled with poisoned (toxic) water at 1000 ppm and 100 ppm levels of poison, respectively, 2826 CSI measurements from WAP signals are recorded. Thus, the total number of data samples which we recorded from these experiments is equal to 5470 records. The procedure of preparing the poisoned water in the glass is as following: preparing 1000 ppm for lead nitrate $Pb(NO_3)_2$, by weight (1gm) of lead nitrate to (1L) of double distilled water (DDW) in graduated volumetric flask. Next, it is injected with (5ml) using syringe to sample water. Thereafter, the bottle is shacked for 5 minutes using shaker type "incubator shaker GFL3031" to complete distribution and mixing of $Pb(NO_3)_2$ with drinking water particles and then putting it into the glass. The same process for 100 ppm repeated to prepare a poisonous water with less $Pb(NO_3)_2$.

## 4- Experiments and Results

This section describes the machine learning algorithms which have been utilised to differentiate clean water from poisoned water. The input for the machine learning classifiers are phase and amplitude features. For each of phase and amplitude, we computed 64 values to keep track of the change in phase and amplitude as a consequence of penetrating the WAP signal by the glass of water. We concatenate the 64 numbers from phase and amplitude to obtain a feature vector of dimension 128 whereby we end up having a dataset of 5470 samples each in dimension 128. We used 4 different classifiers; support vector machines (SVM), k-nearest neighbour, Ensemble learning and Long-Short-Term Memory (LSTM). We implemented these classifiers in MATLAB_R2022b. 5-fold cross validation (5FCV) used to split the data into training and testing and we report average performance metrics for each classifier. We optimise the parameters of each classifier and, in what follows, we give details of parameters and optimisation approaches utilised.

For SVM, we optimised all hyperparameters for each testing fold in the 5FCV scheme. Our experimental investigation reveals that linear and Gaussian kernels are best options for SVM kernel over other kernels. K=1 is best for k-NN after 30 epochs of optimization and 'correlation' is the best distance metric. AdaBoost is the best classifier among ensemble methods for our data after 30 epochs of optimisation trials with 132 as number of splits, number of learners are 466 and learning rate of 0.1241. Finally, the parameters of LSTM is as follows; firstly, we normalise the input using z-score normalisation and then two hidden layers of 200 and 100 nodes respectively. Then Dropout layer and SoftMax layer for classification. The rest of LSTM specification are as follows: the squared gradient decay factor: 0.9, the mini batch size: 150, the shuffling at every-epoch, the initial learning rate is equal to 0.001, the learning rate drop period is 2, maximum Epochs is up to 50, and the L2 regularization is initialized to 5e-4.

The machine learning (ML) algorithms with the above configuration are applied in three different scenarios. With the first scenario, the ML algorithms are only trained and tested on the poisoned water (when 100 ppm for lead nitrate $Pb(NO_3)_2$ and clean water. The classification results of the experiment are shown in Table I. As it can be seen, that all ML algorithms can differentiate the poisoned water from the clean water within the range of 86% up to 89% classification accuracy. The worst classification performance is obtained using k-NN, while the best classification accuracy is obtained by LSTM.

In the second scenario, the same ML algorithms (within the same configuration) are trained and tested on the poisoned water (when 1000 ppm of lead nitrate $Pb(NO_3)_2$ used) and clean water. The classification results show that all tested samples are truly classified into their respective classes, as shown in Table II. This is due to the fact that when the ratio of toxicities is increased, the density of the poisoned water has greater effect on the amplitude and phase features, hence effects on the Wi-F- CSI data.

*Table I: Classification of Clear water and Toxic 100ppm water.*

|         | AUC         | TPR         | TNR         | F1-Score    | Accuracy    |
|---------|-------------|-------------|-------------|-------------|-------------|
| LSTM    | 93.76±1.01  | **89.83±1.11** | 85.77±2.15 | 90.25±0.96 | **89.34±0.78** |
| K-NN    | 85.70±64    | 89.95±0.69  | 80.09±1.28  | 87.89±0.63  | 86.37±0.42  |
| Ensemble| **94.87±0.46** | 87.46±0.77 | **89.99±0.95** | **90.67±0.43** | 88.34±0.59 |
| SVM     | 90.53±4.24  | 87.26±1.26  | 81.76±7.69  | 87.38±3.65  | 87.80±0.84  |

*TABLE II: Classification of Clear water and Toxic 1000ppm water.*

|         | AUC | TPR | TNR | F1-Score | Accuracy |
|---------|-----|-----|-----|----------|----------|
| LSTM    | 100 | 100 | 100 | 100      | 100      |
| K-NN    | 100 | 100 | 100 | 100      | 100      |
| Ensemble| 100 | 100 | 100 | 100      | 100      |
| SVM     | 100 | 100 | 100 | 100      | 100      |

To give general view, with the third scenario, the ML algorithms are applied on the entire collected data samples, i.e. to differentiate the poisoned water (including both 100 ppm and 1000 ppm cases) and the clean water. The results show that the classification accuracy is ranging from 82% by k-NN to 92 % by AdaBoost-Ensemble classifier. In total, the experimental result and the ML algorithms provide promising results to differentiate the poisoned water from the clean water using amplitude and phase features of the received Wi-Fi CSI data. Further, even with the case of the 100 ppm, the achieved 89% of classification accuracy by applying LSTM is a good result with such low-cost solution and without any user intervention, in comparison with other approaches reported in literature to check the quality of drinking water.

*TABLE III: Classification of Clear water, Toxic 1000ppm and 100ppm water.*

|         | AUC         | TPR         | TNR         | F1-Score    | Accuracy    |
|---------|-------------|-------------|-------------|-------------|-------------|
| LSTM    | 96.34±0.5   | 90.82±0.99  | **90.05±1.37** | **90.72±0.77** | 91.59±0.56 |
| K-NN    | 83.68±3.59  | 84.85±1.05  | 80.41±1.34  | 80.91±1.3   | 82.87±0.56  |
| Ensemble| **96.7±0.55** | **93.8±1.23** | 87.75±1.74 | 90.64±1.22 | **92.03±0.64** |
| SVM     | 92.20±2.99  | 86.75±2.78  | 86.16±2.03  | 86.9±1.34   | 86.25±1.67  |

## 5- Conclusions

This article introduces a new method for detecting poisoned water by utilizing wireless sensing technology and machine learning algorithms to distinguish between poisonous and clean water. Proposed method is cost-effective as it does not require additional sensors and can be implemented using the Wi-Fi chipset already built into smartphones. We conducted a set of experiments to show the effectiveness of this method, including comparing model parameters and evaluating it against different machine learning algorithms. The results of these experiments showed that the method was effective, with a classification accuracy of 89% in the worst-case scenario using LSTM and 92% accuracy using Ensemble. We believe that further improvements can be made by incorporating multiple drinking categories of detection and taking into account additional interfering factors, but this falls outside the scope of the current study and will be addressed in future work. This can also be utilized to detect the contamination water from the clean water, whereby the contamination is not a type poisoned water, however, still effect on people's health.

**Author Contributions:**
Conceptualization, **Halgurd S. Maghdid** and **Aras Asaad**
Setting up Hardware (Wi-Fi Chipsets) and collecting data**, Hassan M. Bayram,**
Preparing chemical materials Pb($NO_3$)$_2$, **Salam Ghafour Taher and Kosrat N. Kaka**
Extracting Features, **Halgurd S. Maghdid** and **Safar M. Asaad.**
Implementing machine learning algorithms, **Akar T. Hawre** and **Aras Asaad**
Writing the Manuscript, **Halgurd S. Maghdid, Sheerko R. Hma Salah, Akar T. Hawre, Hassan M. Bayram, Azhin T. Sabir, Aras Asaad, Kosrat N. Kaka, Ladeh S. Abdulrahman, Abdulbasit K. Al-Talabani, Salam Ghafour Taher,** and **Safar M. Asaad.**
Acquisition, **all authors have read and agreed to the published version of the manuscript,**
Funding: This work is supported by Koya University.

**Conflicts of Interest:**
The authors declare no conflict of interest.